\documentclass{hustmedia-paper}

\setcopyright{none}
\usepackage{booktabs}
\usepackage{amsmath}
\usepackage{mathtools}
\usepackage{amsthm}
\usepackage{enumitem}

\usepackage{algorithm}
\usepackage{algpseudocode}
\usepackage[table]{xcolor}
\definecolor{basepurple}{HTML}{bdb2ff}
\usepackage{xcolor}
\definecolor{CMTgreen}{RGB}{0,120,0}
\definecolor{lightgreen}{RGB}{220,255,220}
\definecolor{lightred}{RGB}{255,220,220}

\usepackage[capitalize,noabbrev]{cleveref}

\theoremstyle{plain}
\newtheorem{theorem}{Theorem}[section]
\newtheorem{proposition}[theorem]{Proposition}

\theoremstyle{definition}

\theoremstyle{remark}
\newtheorem{remark}[theorem]{Remark}

\usepackage[textsize=tiny]{todonotes}

\usepackage{xcolor,soul}
\sethlcolor{yellow}

\HMMAuthorLine{
Yunyao Zhang\textsuperscript{1} \textbullet\
Yihao Ai\textsuperscript{2} \textbullet\
Zuocheng Ying\textsuperscript{1} \textbullet\
Qirui Mi\textsuperscript{3} \textbullet\
Junqing Yu\textsuperscript{1} \textbullet\
Wei Yang\textsuperscript{1} \textbullet\
Zikai Song\textsuperscript{1,\ensuremath{\dagger}}
}

\HMMAffiliationLine{
\textsuperscript{1}Huazhong University of Science and Technology;
\textsuperscript{2}National University of Singapore;
\textsuperscript{3}Institute of Automation, Chinese Academy of Sciences
}

\HMMPublicationStatus{Preprint}

\HMMCorrespondingAuthor{
Zikai Song at \HMMEmail{skyesong@hust.edu.cn}.
}

\HMMCorrespondingEmail{skyesong@hust.edu.cn}

\HMMAbstractEmailOn

\HMMAuthorInformation{
Yunyao Zhang, Zuocheng Ying, Junqing Yu, Wei Yang, and Zikai Song ---
Huazhong University of Science and Technology, Wuhan, China;
Yihao Ai --- National University of Singapore, Singapore;
Qirui Mi --- Institute of Automation, Chinese Academy of Sciences, Beijing, China.
}

\HMMDisplayTitle{Coupling Macro Dynamics and Micro States for Long-Horizon Social Simulation}

\HMMProjectWebsite{https://ai4ss.github.io/M2Sim}

\title[Short Paper Title]{Coupling Macro Dynamics and Micro States for Long-Horizon Social Simulation}

\author{Author One}
\affiliation{
  \institution{Huazhong University of Science and Technology}
  \city{Wuhan}
  \country{China}}
\email{author@example.com}

\author{Author Two}
\affiliation{
  \institution{Huazhong University of Science and Technology}
  \city{Wuhan}
  \country{China}}

\begin{abstract}
Social network simulation aims to model collective opinion dynamics in large populations, but existing LLM-based simulators primarily focus on aggregate dynamics and largely ignore individual internal states. 
This limits their ability to capture opinion reversals driven by gradual individual shifts and makes them unreliable over long-horizon simulations.
Unlike existing methods that collapse dynamics into macro-only updates, we propose a social simulation framework, \textbf{M$^2$Sim}, which tightly couples macro-level collective dynamics with micro-level individual states.
We explicitly model micro-level opinion states with a state transition mechanism, coupling \textbf{M}acro-level collective dynamics with \textbf{M}icro-level states. This allows the opinion-state distribution to evolve gradually, rather than triggering instant reactions, enabling the simulator to distinguish states close to or far from switching, capture opinion reversals, and maintain accuracy over long-horizon simulations.
Across real-world events, our M$^2$Sim supports stable simulation of long-horizon social events with up to \textbf{40,000} interactions (compared to $\sim$300 in baseline MF-LLM), while reducing long-horizon KL divergence by \textbf{75.3\%} (1.2490 $\rightarrow$ 0.3089) and reversal KL by \textbf{66.9\%} (1.6425 $\rightarrow$ 0.5434), significantly mitigating the drift observed in MF-LLM.
\end{abstract}


\begin{document}
\maketitle

\begin{figure*}[t]
    \centering
    \includegraphics[width=\linewidth, height=0.45\linewidth]{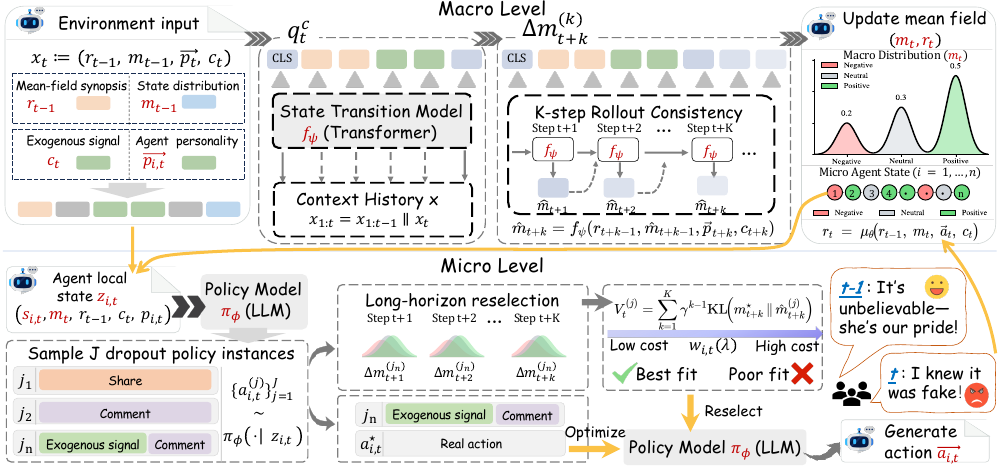}
    \caption{
    \textbf{M$^2$Sim overview.}
    \textbf{(i) Macro level}, an event-level state transition model (Event Transformer) updates the distributional mean field and is trained with a $K$-step rollout consistency objective to improve long-horizon trajectory fidelity.
    \textbf{(ii) Micro level}, a LoRA-tuned policy LLM generates actions conditioned on agent states and the mean-field signals; we sample $J$ dropout-induced policy instances, obtain candidate actions, and score them by a $K$-step mean-field surrogate (distributional KL to future trajectories), yielding soft weights for action reselection and optimizing the policy via a combined auxiliary text loss and discounted long-horizon prediction loss.
    }
    \label{fig2:pipline}
\end{figure*}

\section{Introduction}
Simulating how collective behaviors and social structures emerge is central to understanding social diffusion, mobilization, and opinion formation~\citep{Social-Simulation-overview-2014,social-network-analysis2004development}.
In \textbf{social network simulation (SNS)}~\citep{LLM-Agent-based-simulation-survey-2024large}, macroscopic public outcomes arise from the aggregation of microscopic actions~\citep{diffusion-online-social-networks2017survey}.
This tight micro-macro coupling induces strong nonlinear feedback, such that small initiating groups, delayed evidence, or localized interventions can trigger tipping points and opinion reversals~\citep{5rule-lupeng-2018exploring,peak-time-lupeng2018big}.

Traditional paradigms, including \textit{mechanistic models}~\citep{traditional-simulation-system-dynamics,traditional-simulation-discrete-events}, \textit{empirical and statistical models}~\citep{PSP-2018,shapes-lupeng2019strength}, and \textit{agent-based models}~\citep{first-agent-based-model-schelling-1971,Multiagent-Systems2005}, capture certain macro-micro regularities but rely on static parameters or handcrafted behavioral rules.
As a result, they struggle to represent evolving beliefs, delayed commitment, and feedback-driven opinion transitions at scale, motivating the need for scalable, \emph{state-aware} frameworks~\citep{Micromotives-Macrobehavior-2006}.

Recent studies show that large language models (LLMs)~\citep{Deepseek-r12025deepseek,DeepSeek-OCR2025deepseek} can endow social agents with reasoning, perception, and interaction capabilities, enabling simulations from small-scale communities~\citep{Stanford-town-2023,Stanford1000agents-2024} to multi-scene social systems~\citep{Yulan-onesim2025yulan,Socioverse2025socioverse}.
Frameworks such as \textit{GA-S\textsuperscript{3}}~\citep{GAS32025-ga}, \textit{Oasis}~\citep{Oasis-2025}, and \textit{AgentSociety}~\citep{Agentsociety-2025agentsociety} rely on fully LLM-driven agents to generate rich behaviors.
To improve scalability and macro-level coherence, \textit{MF-LLM}~\citep{MF-LLM2025mf} introduces mean-field modeling into LLM-based simulation, coupling individual actions with aggregated macro signals through iterative feedback.
While \textit{MF-LLM} yields coherent short-term trajectories, it over-compresses individual dynamics into macro-level signals without per-agent latent state modeling.
Consequently, actions are treated as instantaneous reactions rather than state-changing events, which hampers capturing the turning-point timing of opinion reversals and realistic long-horizon collective dynamics.

These gaps highlight three core challenges in social simulation.
\textbf{(C1) Micro-Macro Decoupling}: existing approaches focus on either micro-level individual behaviors or macro-level collective dynamics in isolation, failing to capture the tightly coupled co-evolution of both aspects in real social systems.
\textbf{(C2) Long-Horizon Dynamics Degradation}: achieving \emph{long-horizon} simulation is challenging because single-scale rollouts accumulate errors over time, leading to variance damping, majority lock-in, and unrealistic trajectories under delayed evidence accumulation and gradual opinion transitions.
\textbf{(C3) Unresolved Opinion Reversals}: reversal events impose a stricter requirement on long-horizon simulations. The model must not only sustain long rollouts but also capture timely and significant opinion reversals when influenced by exogenous signals. Without tightly coupled macro-state signals and micro-level opinion states of current participants, simulations exhibit inertia that misses turning points or drift that overshoots, hindering realistic opinion transitions.

\noindent\textbf{Contributions.}
To address challenges (C1-C3), we make the following contributions:


\noindent\textbf{1. M$^2$Sim: A new social dynamic simulation framework that couples macro-level collective dynamics and micro-level agent states.}
We introduce \textbf{M$^2$Sim}, which provides macro-level collective signals through a mean field and generates individual actions conditioned on micro-level opinion states, coupling their evolution through an explicit state-transition mechanism to address the micro-macro decoupling challenge (\textbf{C1}).


\noindent\textbf{2. Coupled state-transition and rollout modeling for long-horizon dynamics and opinion reversals.}
To address \textbf{C2} and \textbf{C3}, M$^2$Sim couples a \emph{macro-level} state transition model with \emph{micro-level} multi-step rollout-based action reselection.
At the macro level, the transition model learns the long-term evolution of the state distribution, reducing drift, variance damping, and majority lock-in.
At the micro level, agents condition on this evolving distribution and use multi-step rollouts to evaluate candidate actions by their future impacts.
This enables agents to adjust their actions in response to exogenous signals, overcoming inertia and capturing the timing and magnitude of opinion reversals.


\noindent \textbf{3. Empirical Validation.}  
Experiments on diverse real-world social events demonstrate that \textbf{M$^2$Sim} consistently outperforms existing methods: it significantly reduces long-horizon KL divergence by \textbf{75.3\%} and reversal KL by \textbf{66.9\%}, while mitigating the drift commonly observed in \textit{MF-LLM}. Additionally, \textbf{M$^2$Sim} maintains the scalability advantages in long-horizon scenes, enabling stable simulations even with up to \textbf{40,000} interactions (vs.\ $\sim$300).

\section{Background and Motivation}
We formalize collective dynamics as a state-aware decision process (§\ref{sec3.1:problem}) and show that omitting micro-level states induces self-reinforcing rollouts that miss opinion reversals (Prop.~\ref{prop:self_strengthening}).

\subsection{Problem and Notation Definition}
\label{sec3.1:problem}
We consider a social group of $N$ agents and simulate its collective decision-making over $[0,T]$.
All agents share a discrete action space $\mathcal{A}=\{a_1,a_2,\ldots,a_n\}$ (e.g., comment, share) and state space $\mathcal{S}=\{s_1,s_2,\ldots,s_m\}$.
The action and state of agent $i$ at time $t$ are $a_{i,t}$ and $s_{i,t}$, with aggregated vectors
$\vec{a}_t=[a_{1,t},\ldots,a_{N,t}]$ and $\vec{s}_t=[s_{1,t},\ldots,s_{N,t}]$.
Each agent has a personality descriptor $p_{i,t}$, collected as $\vec{p}_t=[p_{1,t},\ldots,p_{N,t}]$, which modulates its decision behavior.
Although $\mathcal{S}$ and $\mathcal{A}$ are written as finite sets for convenience, they correspond in practice to latent or semantic categories derived from open-ended interactions.
To characterize macroscopic dynamics, we introduce a mean-field representation $m_t\in\Delta^{m-1}$, defined as the macro state distribution over $\mathcal{S}$ at time $t$, where
\[
\Delta^{m-1}
=
\Bigl\{x\in\mathbb{R}^m_{\ge 0}:\sum_{s\in\mathcal{S}} x(s)=1\Bigr\}
\]
denotes the probability simplex over $\mathcal{S}$. The evolution of $m_t$ captures how decentralized agent decisions give rise to collective behavioral trends, which we also refer to as the \emph{public mood} in the context of social group simulation. The system is driven by exogenous sianals $c_t$, which are generated endogenously within the simulator by designated information-source agents (e.g., official accounts or media reporters) at step $t$ and then broadcast to the public.
These externally grounded messages influence agent states and actions and thereby shape the joint evolution of individual actions and the mean-field trajectory $\{m_t\}_{t=0}^T$, capturing both short-term responses and long-term collective dynamics in social propagation. In addition, we maintain a textual mean-field synopsis $r_t\in\mathbb{T}$, which provides a natural-language summary of recent collective trajectories and exogenous signals; together, $(r_t,m_t)$ constitute the macros. 

\subsection{Dynamics of Previous Method}
The existing simulation dynamics evolve by two parameterized models. First, the action model $\pi_{\phi}(a_{i,t}|r_t, p_{i,t}, c_t)$, it samples the actions of $i$-th agent based on the current public mood (mean-field) $r_t$, the personality of the agent $p_{i,t}$ at time $t$. Second a mean-field summarizer $\mu_{\theta}(r_{t+1}|r_t, \mathbf{a}_t)$. It summarize the next public mood ($r_{t+1}$) based on the current actions $\mathbf{a}_t$ and the current public mood ($r_{t}$). The model iteratively updates $\mathbf{a}_t$ and $r_t$, which stops until convergence or the reaching the terminal time $T$. It is obvious that this dynamic is state-ignored.

\subsection{Limitations of State-ignored Dynamics.}
A key parameter that tracks evolution in $\{m_t\}_{t=0}^T$ is $f(m) := \mathbb{E}[m_{t+1}|m_t]$, which indicates the expected $m_{t+1}$ given the previous $m_t$. The $\delta m(s):=\mathbb{E}[m_{t+1}(s)|m_t] -  m(s)$ indicates the expected changes of the mean field over a specific state $s$. We find the following proposition w.r.t. the majority state $s^* =  \arg\max_{s \in \mathcal{S}} m(s)$ under very mild assumptions: (1). The mean-field summarizer $\mu_{\theta}$ faithfully summarize the $m_t$ with a bounded error $\epsilon$ for some very small $\epsilon \ge 0$, i.e., 
\begin{equation}
\mathbb{E}[m_{t+1}(s)|m_t, \mathbf{a}_t] \ge \hat{q}(\mathbf{a}_t)(s)-\epsilon,
\end{equation} where the $\hat{q}(\mathbf{a}_t)$ is the empirical distribution of $\mathcal{S}$ yet projected from actions $\mathcal{A}$; (2). The agents, "on average", align slightly toward the current majority of the public mood, i.e.,
\begin{equation}
\mathbb{E} [\hat{q}(\mathbf{a}_t)(s^*)|m_t=m] \ge m(s^*) + \eta, 
\end{equation} for some $\eta > 0$.

\begin{proposition} [Self-strengthening of state-ignore dynamics] After the last exogenous signals $c_t$, the $\delta m(s^*) \ge \eta - \epsilon \ge0, \text{for any } \eta \ge \epsilon. $
\label{prop:self_strengthening}
\end{proposition}
\begin{remark}
In practice, the aggregation error $\epsilon$ is empirically small because $\mu_{\theta}$ is prompted and trained to produce a faithful summary (sometimes even deterministic). Consequently, the one-step expected majority drift $\delta m(s^*) \ge 0$ in the scenario that a clear majority of public mood exists. This implies a systematic tendency to reinforce the current majority after the last exogenous signal. Though, this does not rule out reversals of personal or public mood, yet reversals become increasingly rare and typically require atypical realizations of the action-sampling noise.
\end{remark}

\section{Methodology}
We now introduce the \textbf{M$^2$Sim} framework (§\ref{sec3.2:M$^2$Sim}) to couple macro-level states with micro-level agent states (Fig.~\ref{fig2:pipline}).
Then, we present \textbf{LCT} (§\ref{sec3.3:LCT-Tune}), a trajectory-aware algorithm for realistic reversal and long-horizon simulation.
The overall procedure is summarized in Algorithm~\ref{alg:M$^2$Sim}.

\subsection{M$^2$Sim Framework}
\label{sec3.2:M$^2$Sim}
To model public mood dynamics at scale, we introduce \textbf{M$^2$Sim}, where the mean field captures the macro-level state distribution and a state-transition mechanism models micro-level agent dynamics.

\subsubsection*{Macro-Level Mean Field}

Recall from §\ref{sec3.1:problem} that the macroscopic mean field at time $t$ is given by the pair
\[
(r_t, m_t),
\quad
r_t\in\mathbb{T},\;
m_t\in\Delta^{m-1},
\]
where $\mathbb{T}$ denotes the space of textual summaries and $\Delta^{m-1}$ is the probability simplex over the state space $\mathcal{S}$. We view these as two synchronized channels:

\begin{itemize}[leftmargin=*]  
    \item \textbf{Textual synopsis $r_t$:} a free-form natural-language summary of recent trajectories, in the same spirit as the memory string in \textit{MF-LLM}. It aggregates salient information about past micro-level interactions and exogenous signals into a compact macro-level description.
    \item \textbf{Distributional synopsis $m_t$:} a macro-level distribution over $\mathcal{S}$ that retains accumulated opinion-state information from past participation and incorporates updates from current participants. It explicitly tracks the relative proportions of latent opinion states. When \(|\mathcal{S}|\) is large, semantically related states may be merged into coarse buckets to keep $m_t$ low-dimensional.
\end{itemize}

\paragraph{Recursive update.} 
At each time step $t\ge 1$, the macroscopic signal $(r_{t-1},m_{t-1})$ and exogenous sianal $c_t$ drive the evolution of internal states and the public mood.

First, $f_{\psi}$ generates current participants' states and updates the distributional mean field:
\begin{equation}
\label{eq:mf_state_update}
\bigl(\vec{s}_t,\, m_t\bigr)
\;=\;
f_{\psi}\!\bigl(m_{t-1},\; r_{t-1},\; \vec{p}_t,\; c_t\bigr),
m_t\in\Delta^{m-1},
\end{equation}
where $n_t\in\{0,\ldots,N\}$ denotes the number of active participants at time $t$, and
$\vec{s}_t=[s_{1,t},\ldots,s_{n_t,t}]$ and $\vec{p}_t=[p_{1,t},\ldots,p_{n_t,t}]$ are their opinion states and personality descriptors, respectively.

After actions $\vec{a}_t=[a_{1,t},\ldots,a_{n_t,t}]$ have been realised at time $t$, the textual synopsis is updated by the \textit{MF-LLM} summariser $\mu_{\theta}$:
\begin{equation}
\label{eq:mf_text_update}
r_t
\;=\;
\mu_{\theta}\!\bigl(r_{t-1},\; m_t,\; \vec{a}_t,\; c_t\bigr).
\end{equation}

In the degenerate case where internal states and the distributional mean field are ignored, the effective state collapses to the textual synopsis $r_t$.
Concretely, dropping $\vec{s}_t$ and $m_t$, we update $r_t$ only from $(r_{t-1},\vec{a}_t,c_t)$ and condition each participant's action on $(r_{t-1},p_{i,t},c_t)$.
This recursion coincides with macro-only two-LLM mean-field simulators, where a single textual mean field both conditions agent behaviour and summarizes collective trajectories.

\paragraph{Warm-up phase.}
Following \textit{MF-LLM}, we employ a short warm-up phase of length $t_{\mathrm{warm}}$ in which the textual synopsis $r_t$ is updated using ground-truth trajectories $\mathbf{a}_t^{\ast}$ and corresponding states, providing a stable initial macro description. Unlike the baseline, we \emph{also} initialize the distributional channel from empirical data: $m_0$ is set to the historical state frequency over $\mathcal{S}$ rather than a uniform prior. This ensures that the mean field starts from a realistic macro configuration before simulated dynamics take over.

\subsubsection*{Micro-Level Agents}

Given the macroscopic mean field $(r_{t-1},m_t)$, each agent acts based on its opinion state, personality, and macro-level information.
For agent $i$, we denote this combined decision context by
\(z_{i,t} = (s_{i,t}, r_{t-1}, m_t, c_t, p_{i,t}) \in \mathcal{Z}\).

All agents share the same action space $\mathcal{A}$ and follow a stochastic policy
\begin{equation}
\pi_{\phi}(a_{i,t}\mid z_{i,t})
\;=\;
\pi_{\phi}\bigl(a_{i,t}\mid s_{i,t},\, r_{t-1},\, m_t,\, c_t,\, p_{i,t}\bigr)
\label{eq:agent_policy}
\end{equation}
which maps the current decision context to a distribution over actions, capturing that decisions are shaped jointly by micro-level opinion and personality $(s_{i,t},p_{i,t})$ and macro-level information $(r_{t-1},m_t,c_t)$.

Temporal evolution is not modelled via identity-specific kernels, but through the macro-level transition encoded in the mean field.
At each time step, the state model $f_{\psi}$ generates current participants' states and updates $m_t$ from the previous macroscopic mean field, participant personalities, and the exogenous signal, retaining historical opinion-state information.
Current participants act according to $\pi_{\phi}$ in~\eqref{eq:agent_policy}.
This recursive coupling between $(r_{t-1},m_t,c_t)$ and agent policies links micro-level actions with macro-level dynamics; in the degenerate case where internal states and the distributional mean field are ignored, actions depend only on $(r_{t-1},p_{i,t},c_t)$ and textual updates on $(r_{t-1},\vec{a}_t,c_t)$, recovering prior macro-only two-LLM simulators as a special case.

\paragraph{Environment transition dynamics.}
We summarize the updates within step $t$ and the preparation for step $t+1$ through an environment transition.
In M$^2$Sim, this transition jointly describes updates to the distributional channel, the textual channel, the exogenous signal, and the active agent pool:
\begin{equation}
\label{eq:env_transition}
\bigl(m_t,\; r_t,\; \vec{p}_{t+1},\; c_{t+1}\bigr)
\;=\;
\mathcal{T}\!\bigl(\vec{p}_t,\; m_{t-1},\; r_{t-1},\; \vec{a}_t,\; c_t\bigr).
\end{equation}

\subsection{Long-Horizon Consistency Training}
\label{sec3.3:LCT-Tune}
While \textit{MF-LLM} (the textual synopsis $r_t$) provides a scalable macroscopic summary, its fidelity can be improved from both macro and micro perspectives. We propose \textbf{LCT} to optimize (1) \emph{macro-level:} the state transition model, to extract predictive macroscopic signals; and (2) \emph{micro-level:} the policy model, to generate behaviorally realistic actions conditioned on these signals.

\subsubsection*{State Transition Model}

We model macro dynamics as learning a conditional law over \emph{mean-field sequences}.
Let $m_t\in\Delta^{m-1}$ denote the distributional mean field at step $t$, and define the event-level context
\[
x_t := (r_{t-1},\, m_{t-1},\, \vec{p}_t,\, c_t).
\]
We instantiate the mean-field state transition model $f_{\psi}$ as a temporal sequence Transformer that maps the context history to a prediction of the current mean field:
\[
\hat m_t \;=\; f_{\psi}\!\bigl(x_{1:t}\bigr)\in\Delta^{m-1},
\]
where self-attention provides a flexible mechanism to capture delayed evidence accumulation by selectively attending to relevant past contexts.

Given empirical trajectories $\{m_t^{\star}\}_{t=1}^{T}$, we train $f_{\psi}$ by matching the \emph{entire} predicted sequence to the observed sequence using a KL-based sequence loss:
\begin{equation}
\label{eq:seq_kl}
\mathcal{L}_{\mathrm{seq}}(\psi)
\;=\;
\mathbb{E}\Bigg[\sum_{t=1}^{T}
\mathrm{KL}\!\left(m_t^{\star}\,\|\,\hat m_t\right)\Bigg],
\hat m_t = f_{\psi}(x_{1:t}).
\end{equation}

Since simulation quality depends on multi-step evolution, we further impose a \emph{rollout consistency} objective over a short horizon $K$.
Starting from the empirical state $m_t^{\star}$, we generate a $K$-step rollout by recursively feeding the model with its own predictions:
\begin{equation}
\label{eq:rollout_recursion}
\hat m_{t+k}
=
f_{\psi}\!\bigl(r_{t+k-1},\,\hat m_{t+k-1},\,\vec{p}_{t+k},\,c_{t+k}\bigr),
\end{equation}
where the rollout is initialized at $\hat m_t=m_t^{\star}$ and recursively applied for $k=1,\ldots,K$.

We then penalize the discrepancy between the rolled-out predictions and the observed future distributions:
\begin{equation}
\label{eq:rollout_kl}
\mathcal{L}_{\mathrm{roll}}(\psi)
\;=\;
\mathbb{E}\Bigg[\sum_{t=1}^{T-K}\;\sum_{k=1}^{K}
\mathrm{KL}\!\left(m_{t+k}^{\star}\,\|\,\hat m_{t+k}^{(k)}\right)\Bigg].
\end{equation}
The final training objective combines sequence fitting and rollout consistency,
\begin{equation}
\label{eq:trans_final}
\mathcal{L}_{\mathrm{trans}}(\psi)
\;=\;
\mathcal{L}_{\mathrm{seq}}(\psi) \;+\; \alpha\,\mathcal{L}_{\mathrm{roll}}(\psi),
\end{equation}
encouraging $f_{\psi}$ to both match observed trajectories and remain stable under short self-conditioned rollouts, which is crucial for reproducing realistic delayed-commitment and opinion reversal.

\subsubsection*{Policy Model Optimization}

The policy model $\pi_{\phi}$ governs microscopic decision-making conditioned on private states and the macroscopic mean-field signal.
We adopt a factorized structure over the active agent set at step $t$,
\begin{equation}
\pi_{\phi}(\vec{a}_t \mid \vec{z}_t)
\;=\;
\prod_{i=1}^{n_t}\pi_{\phi}(a_{i,t}\mid z_{i,t}),
\label{eq:policy_factorized}
\end{equation}
which is scalable while allowing heterogeneous behaviors through agent-specific inputs.

\paragraph{Long-horizon action reselection with dropout policy sampling.}
Direct RL for LLM policies is often bottlenecked by action sampling, since exploring diverse behaviors requires many autoregressive rollouts.
To enable efficient long-horizon optimization, we shift exploration from token trajectories to \emph{latent policy instances} by sampling dropout subnetworks.
Concretely, we sample a dropout variable $\lambda$ (implemented by in-place dropout in the forward pass), which induces a conditional policy $\pi_\phi(\cdot\mid\lambda)$.
We then score each $\lambda$ using a long-horizon mean-field surrogate computed via the state transition model, avoiding explicit text rollouts.
The full derivation is provided in Appendix~\ref{app:B}.

\paragraph{Training objective.}
We define a long-horizon cost for a candidate policy instance as
\begin{equation}
V_\phi(\vec{a},\lambda)=\sum_{k=1}^{K}\gamma^{k-1}\mathrm{KL}\!\left(m_{t+k}^{\star}\,\|\,\hat m_{t+k}(\pi_\phi(\vec{a}),\lambda)\right),
\end{equation}
where $K$ is the rollout horizon and $\gamma\in(0,1]$ is the discount factor.
We optimize a weighted prediction loss $\mathcal{L}_{\rm pred}$ using soft weights over dropout samples:
\begin{align}
\mathcal{L}_{\rm pred}
&= \sum_{t=1}^{T}\sum_{i=1}^{n_t}
\mathbb{E}_{\lambda \sim p_{\rm trivial}(\lambda)}
\Bigl[
w_{i,t}(\lambda)\, V_\phi(\vec{a}_{i,t},\lambda)
\Bigr],
\end{align}
where $\vec{a}_{i,t}\sim \pi_\phi(\cdot\mid\lambda)$ denotes the candidate action of agent $i$ at step $t$ under dropout sample $\lambda$.
Here $p_{\rm trivial}(\lambda)$ is a tractable reference distribution for dropout sampling, and $w_{i,t}(\lambda)$ is the soft weight over sampled $\lambda$, computed as a softmax on the corresponding long-horizon cost.

To stabilize training, we add an auxiliary text-supervision loss
{\footnotesize
\begin{equation}
\mathcal{L}_{\mathrm{text}}(\phi)
=
-\mathbb{E}_{\lambda \sim p_{\rm trivial}(\lambda)}\!\left[\sum_{t=1}^{T}\sum_{i=1}^{n_t}
\log \pi_{\phi}\!\bigl(a_{i,t}^{\star}\mid \lambda\bigr)\right].
\end{equation}
}
The final objective is
\begin{equation}
\mathcal{L}_{\mathrm{total}}(\phi)
=\mathcal{L}_{\mathrm{pred}}(\phi) +
\alpha\,\mathcal{L}_{\mathrm{text}}(\phi).
\end{equation}

\algtext*{EndIf}
\begin{algorithm}[t]
\caption{\textbf{M$^2$Sim Simulation Procedure}}
\label{alg:M$^2$Sim}
\small
\begin{algorithmic}[1]
\Require Active agent pool size $n_t\in N$, horizon $T$, warm-up length $t_{\mathrm{warm}}$;
state transition model $f_{\psi}$, \textit{MF-LLM} summariser $\mu_{\theta}$, agent policy $\pi_{\phi}$;
exogenous sianals $\{c_t\}_{t=1}^{T}$
\State \textbf{Initialize:} $(r_0,m_0)$, active pool descriptors $\vec{p}_1$
\For{$t=1$ \textbf{to} $T$}
    \State \textcolor{CMTgreen}{// Update distributional channel (and latent states)}
    \State $(\vec{s}_t,m_t)\gets f_{\psi}(m_{t-1},r_{t-1},\vec{p}_t,c_t)$  \Comment{Eq.~\eqref{eq:mf_state_update}} 
    
    \If{$t \le t_{\mathrm{warm}}$}
        \State \textcolor{CMTgreen}{// Warm-up: use background actions and states}
        \State $\vec{a}_t \gets \vec{a}^{\ast}_t$; \quad $m_t \gets m_t^{\ast}$
    \Else
        \State \textcolor{CMTgreen}{// Simulation: sample actions by policy}
        \ForAll{$i\in\{1,\ldots,n_t\}$} \Comment{active subset at step $t$}
            \State $z_{i,t}\gets (s_{i,t},r_{t-1},m_t,c_t,p_{i,t})$
            \State $a_{i,t}\sim \pi_{\phi}(\cdot\mid z_{i,t})$ \Comment{Eq.~\eqref{eq:agent_policy}}
        \EndFor
        \State $\vec{a}_t\gets [a_{1,t},\ldots,a_{n_t,t}]$
    \EndIf
    
    \State \textcolor{CMTgreen}{// Update textual channel}
    \State $r_t \gets \mu_{\theta}(r_{t-1},m_t,\vec{a}_t,c_t)$ \Comment{Eq.~\eqref{eq:mf_text_update}}
    
    \State \textcolor{CMTgreen}{// Environment transition (agent pool refresh)}
    \State $\vec{p}_{t+1}\gets \textsc{Refresh}(\vec{p}_t;\,t)$ \Comment{induces Eq.~\eqref{eq:env_transition}}
\EndFor
\State \textbf{Output:} Sequence of simulated actions $\{\vec{a}_t\}_{t=0}^{T}$
\end{algorithmic}
\end{algorithm}

\begin{table*}[t]
\centering
\caption{\textbf{Comparison across simulation settings.}
We report distributional and classification metrics under
\textsc{short-horizon}, \textsc{long-horizon}, and \textsc{reversal}
simulations. Lower is better ($\downarrow$) for all metrics except
Micro/Macro-F1 ($\uparrow$).
Values denote mean with improvement rate (\%) relative to
\textbf{Direct LLM}.
\textbf{Bold} indicates the best and
\underline{underline} indicates the second best.}

\begin{tabular*}{\textwidth}{@{\extracolsep{\fill}}lcccccc@{}}
\toprule
\textbf{Method} & \textbf{KL Div.}$\downarrow$ & \textbf{Wass. Dist.}$\downarrow$ & \textbf{DTW}$\downarrow$ & \textbf{Macro F1}$\uparrow$ & \textbf{Micro F1}$\uparrow$ & \textbf{NLL Loss}$\downarrow$ \\
\midrule

\multicolumn{7}{l}{\textsc{Short-Horizon} (\textsc{Default Steps})} \\
\textbf{Direct LLM} &
0.1101 & 0.1661 & 0.1578 & 0.5805 & 0.6953 & 4.0674 \\
\textbf{Social Retrieval} &
\cellcolor{basepurple!5}0.1051{\scriptsize~(4.54\%)} &
\cellcolor{basepurple!5}0.1578{\scriptsize~(5.00\%)} &
\cellcolor{basepurple!5}0.1500{\scriptsize~(4.94\%)} &
\cellcolor{basepurple!1}0.5829{\scriptsize~(0.41\%)} &
0.6905{\scriptsize~(-0.69\%)} &
\cellcolor{basepurple!3}3.9532{\scriptsize~(2.81\%)} \\
\textbf{MF-LLM} &
\underline{\cellcolor{basepurple!55}0.0492{\scriptsize~(55.31\%)}} &
\underline{\cellcolor{basepurple!36}0.1062{\scriptsize~(36.06\%)}} &
\textbf{\cellcolor{basepurple!40}0.0944{\scriptsize~(40.18\%)}} &
\underline{\cellcolor{basepurple!1}0.5861{\scriptsize~(0.96\%)}} &
\underline{\cellcolor{basepurple!1}0.6975{\scriptsize~(0.32\%)}} &
\underline{\cellcolor{basepurple!3}3.9336{\scriptsize~(3.29\%)}} \\
\textbf{M$^2$Sim (Ours)} &
\textbf{\cellcolor{basepurple!59}0.0453{\scriptsize~(58.86\%)}} &
\textbf{\cellcolor{basepurple!39}0.1006{\scriptsize~(39.43\%)}} &
\underline{\cellcolor{basepurple!37}0.0995{\scriptsize~(36.95\%)}} &
\textbf{\cellcolor{basepurple!2}0.5897{\scriptsize~(1.58\%)}} &
\textbf{\cellcolor{basepurple!2}0.7082{\scriptsize~(1.86\%)}} &
\textbf{\cellcolor{basepurple!4}3.9156{\scriptsize~(3.73\%)}} \\
\midrule

\multicolumn{7}{l}{\textsc{Long-Horizon} (\textsc{Full Steps})} \\
\textbf{Direct LLM} &
1.8300 & 0.3746 & 0.4003 & 0.3922 & 0.5621 & 4.6141 \\
\textbf{Social Retrieval} &
\cellcolor{basepurple!10}1.6554{\scriptsize~(9.54\%)} &
\cellcolor{basepurple!6}0.3515{\scriptsize~(6.17\%)} &
\cellcolor{basepurple!5}0.3816{\scriptsize~(4.67\%)} &
\cellcolor{basepurple!2}0.4014{\scriptsize~(2.35\%)} &
\cellcolor{basepurple!6}0.5962{\scriptsize~(6.07\%)} &
\cellcolor{basepurple!2}4.5294{\scriptsize~(1.84\%)} \\
\textbf{MF-LLM} &
\underline{\cellcolor{basepurple!32}1.2490{\scriptsize~(31.75\%)}} &
\underline{\cellcolor{basepurple!13}0.3251{\scriptsize~(13.21\%)}} &
\underline{\cellcolor{basepurple!28}0.2886{\scriptsize~(27.91\%)}} &
\underline{\cellcolor{basepurple!18}0.4643{\scriptsize~(18.38\%)}} &
\underline{\cellcolor{basepurple!9}0.6153{\scriptsize~(9.46\%)}} &
\underline{\cellcolor{basepurple!15}3.9198{\scriptsize~(15.05\%)}} \\
\textbf{M$^2$Sim (Ours)} &
\textbf{\cellcolor{basepurple!83}0.3089{\scriptsize~(83.12\%)}} &
\textbf{\cellcolor{basepurple!53}0.1773{\scriptsize~(52.67\%)}} &
\textbf{\cellcolor{basepurple!58}0.1666{\scriptsize~(58.38\%)}} &
\textbf{\cellcolor{basepurple!29}0.4805{\scriptsize~(22.51\%)}} &
\textbf{\cellcolor{basepurple!14}0.6393{\scriptsize~(13.73\%)}} &
\textbf{\cellcolor{basepurple!18}3.7655{\scriptsize~(18.39\%)}} \\
\midrule

\multicolumn{7}{l}{\textsc{Reversal} (\textsc{Full Steps})} \\
\textbf{Direct LLM} &
5.4883 & 0.4535 & 0.4377 & 0.3531 & 0.4837 & 4.6469 \\
\textbf{Social Retrieval} &
\cellcolor{basepurple!35}3.5539{\scriptsize~(35.26\%)} &
\cellcolor{basepurple!2}0.4432{\scriptsize~(2.27\%)} &
\cellcolor{basepurple!9}0.3978{\scriptsize~(9.12\%)} &
\cellcolor{basepurple!5}0.3690{\scriptsize~(4.50\%)} &
\cellcolor{basepurple!4}0.5017{\scriptsize~(3.72\%)} &
\cellcolor{basepurple!10}4.2002{\scriptsize~(9.61\%)} \\
\textbf{MF-LLM} &
\underline{\cellcolor{basepurple!70}1.6425{\scriptsize~(70.08\%)}} &
\underline{\cellcolor{basepurple!39}0.2763{\scriptsize~(39.06\%)}} &
\underline{\cellcolor{basepurple!45}0.2425{\scriptsize~(44.59\%)}} &
\underline{\cellcolor{basepurple!18}0.4158{\scriptsize~(17.76\%)}} &
\underline{\cellcolor{basepurple!18}0.5721{\scriptsize~(18.27\%)}} &
\underline{\cellcolor{basepurple!16}3.8841{\scriptsize~(16.41\%)}} \\
\textbf{M$^2$Sim (Ours)} &
\textbf{\cellcolor{basepurple!90}0.5434{\scriptsize~(90.10\%)}} &
\textbf{\cellcolor{basepurple!53}0.2127{\scriptsize~(53.10\%)}} &
\textbf{\cellcolor{basepurple!55}0.1986{\scriptsize~(54.63\%)}} &
\textbf{\cellcolor{basepurple!28}0.4533{\scriptsize~(28.38\%)}} &
\textbf{\cellcolor{basepurple!25}0.6065{\scriptsize~(25.39\%)}} &
\textbf{\cellcolor{basepurple!17}3.8384{\scriptsize~(17.40\%)}} \\
\bottomrule
\end{tabular*}
\label{tab1:comparison}
\end{table*}

\section{Experiment}
\subsection{Settings}
\textit{Details are in the Appendix~\ref{app:C}; here we list what is used.}

\noindent\textbf{Model.}
Our framework contains two trainable components.
\textbf{(1) State transition model.} We use an event-level causal Transformer to predict the macro state distribution $m_t$ from historical mean-field signals.
\textbf{(2) Policy model.} Following \textit{MF-LLM}, we adopt \textbf{Qwen2-1.5B} as the backbone and fine-tune it with LoRA, attaching a lightweight $K$-step predictor to support long-horizon action selection.
\textbf{Training.} We train for $1$ epoch with learning rate $1\times10^{-5}$ (other detailed settings in Appendix~\ref{app:C.1 training}).

\noindent\textbf{Dataset.}
We follow \textit{MF-LLM}~\citep{MF-LLM2025mf} and use the WEIBO corpus~\citep{weibo-2016}.
To stress-test reversal and delayed-commitment dynamics, we additionally crawl reversal events from Weibo and Douyin and convert them into the same \textit{MF-LLM} event-centric format.
For WEIBO, we select test events with trajectory length $>1000$ and evaluate both (i) the default rollout horizon and (ii) a long-horizon rollout over the full trajectory.
For the reversal set we simulate full trajectories.

\noindent\textbf{Evaluation Metrics.}
We adopt a micro-to-macro evaluation protocol.
(i) Micro level, we annotate each individual action into one of $8$ evaluation dimensions.
(ii) Macro level, we compute the action distribution over $n_t$ actions at each timestep $t$ and report
(1) \textbf{KL Divergence}, (2) \textbf{Wasserstein Distance}, (3) \textbf{DTW}, (4) \textbf{NLL}, (5) \textbf{Macro-F1}, and (6) \textbf{Micro-F1}.
Details are provided in Appendix~\ref{appC.2:detailed setup}.

\noindent\textbf{Baselines.}
We compare M$^2$Sim with representative LLM-based baselines:
(1) \textit{Direct LLM~\citep{s3-2023,Stanford-town-2023}}, which conditions a vanilla LLM only on the individual profile and event topic/context;
(2) \textit{Social Retrieval~\citep{Agentsociety-2025agentsociety,Socioverse2025socioverse}}, which augments Direct LLM with retrieved peer responses (the $k$ most recent and $k$ most popular comments);
(3) \textit{MF-LLM~\citep{MF-LLM2025mf}}, the two-LLM mean-field simulator with a textual synopsis for summarization.

\subsection{Comparison with Baselines}

The main results are summarized in Table~\ref{tab1:comparison} (full in Appendix~\ref{appD.1:full result}); we make three observations.

\textbf{(1) Strong short-horizon performance.}
Under \textsc{short-horizon} (DEFAULT STEPS) simulation, \textit{MF-LLM} already achieves strong distributional and classification accuracy, indicating that a textual mean field is sufficient for capturing near-term macro trends.
M$^2$Sim yields only a modest but consistent improvement on top of MF-LLM, suggesting that explicit state distributions and stateful dynamics mainly contribute beyond the short-horizon regime.

\textbf{(2) Long-horizon robustness.}
Under \textsc{long-horizon} (FULL STEPS) simulation, all baselines, including \textit{Direct LLM}, \textit{Social Retrieval}, and \textit{MF-LLM}, degrade, reflecting compounding errors when the rollout extends beyond the default window.
While \textit{MF-LLM} improves over \textit{Direct LLM} on distributional metrics (e.g., KL $1.8300 \rightarrow 1.2490$, \textbf{31.7\%} reduction; DTW $0.4003 \rightarrow 0.2886$, \textbf{27.9\%} reduction), its long-rollout drift remains pronounced.
In contrast, M$^2$Sim remains stable and improves over \textit{MF-LLM} (KL $1.2490 \rightarrow 0.3089$, \textbf{75.3\%} reduction; DTW $0.2886 \rightarrow 0.1666$, \textbf{42.3\%} reduction), demonstrating stronger long-horizon consistency in tracking collective trajectories.

\textbf{(3) Reversal opinion dynamics.}
The reversal setting is the most challenging because it requires models to capture non-monotonic trend changes and delayed commitment.
\textit{MF-LLM} improves over \textsc{Social Retrieval} on trajectory metrics (e.g., KL $3.5539 \rightarrow 1.6425$, \textbf{53.8\%} reduction), yet its reversal tracking remains noticeably misaligned.
M$^2$Sim further strengthens reversal fidelity (KL $1.6425 \rightarrow 0.5434$, \textbf{66.9\%} reduction; DTW $0.2425 \rightarrow 0.1986$, \textbf{18.1\%} reduction) and improves classification accuracy (e.g., Micro-F1 $0.5721 \rightarrow 0.6065$, \textbf{6.0\%} increase), indicating that explicit distributional signals and long-horizon consistency training better support reversal dynamics over extended timelines.

\newcommand{\better}[2]{\fcolorbox{white}{lightgreen}{#1}\,{\color{green!60!black}(+#2)}}
\newcommand{\worse}[2]{\fcolorbox{white}{lightred}{#1}\,{\color{red!70!black}(-#2)}}
\newcommand{\same}[1]{#1}

\begin{table*}[t]
\centering
\caption{\textbf{Ablation study on M$^2$Sim.} We evaluate the contribution of LCT-State, LCT-Policy and Sampling under \textsc{short-horizon}, \textsc{long-horizon}, and \textsc{reversal} simulations.
\fcolorbox{white}{lightgreen}{Green} = improvement; 
\fcolorbox{white}{lightred}{Red} = degradation; 
\textbf{Bold} = best.}
\setlength{\tabcolsep}{1pt}
\begin{tabular}{lcccccc}
\toprule
\textbf{Method} & \textbf{KL Div.}$\downarrow$ & \textbf{Wass. Dist.}$\downarrow$ & \textbf{DTW}$\downarrow$ & \textbf{Macro F1}$\uparrow$ & \textbf{Micro F1}$\uparrow$ & \textbf{NLL Loss}$\downarrow$ \\
\midrule

\multicolumn{7}{l}{\textsc{Short-Horizon} (\textsc{Default Steps})} \\
\textbf{M$^2$Sim (Ours)} &
0.0453 & \textbf{0.1006} & 0.0995 & 0.5897 & \textbf{0.7082} & \textbf{3.9156} \\
\textbf{w/o LCT-State} &
\better{0.0449}{0.88\%} & \worse{0.1120}{11.33\%} & \better{\textbf{0.0802}}{19.40\%} &
\worse{0.5825}{1.22\%} & \worse{0.6828}{3.59\%} & \worse{4.0064}{2.32\%} \\
\textbf{w/o LCT-Policy} &
\worse{0.0491}{8.39\%} & \worse{0.1051}{4.47\%} & \better{0.0856}{13.97\%} &
\worse{0.5809}{1.49\%} & \worse{0.6962}{1.69\%} & \worse{3.9403}{0.63\%} \\
\textbf{w/o Sampling} &
\worse{0.0649}{43.27\%} & \worse{0.1115}{10.83\%} & \worse{0.1081}{8.64\%} &
\better{\textbf{0.5925}}{0.47\%} & \worse{0.7014}{0.96\%} & \worse{3.9206}{0.13\%} \\
\midrule

\multicolumn{7}{l}{\textsc{Long-Horizon} (\textsc{Full Steps})} \\
\textbf{M$^2$Sim (Ours)} &
\textbf{0.3089} & \textbf{0.1773} & \textbf{0.1666} & \textbf{0.4805} & \textbf{0.6393} & \textbf{3.7655} \\
\textbf{w/o LCT-State} &
\worse{0.8770}{183.91\%} & \worse{0.2123}{19.74\%} & \worse{0.2104}{26.29\%} &
\worse{0.4725}{1.66\%} & \worse{0.6187}{3.22\%} & \worse{3.9414}{4.67\%} \\
\textbf{w/o LCT-Policy} &
\worse{0.5872}{90.09\%} & \worse{0.2028}{14.38\%} & \worse{0.1730}{3.84\%} &
\worse{0.4314}{10.22\%} & \worse{0.5765}{9.82\%} & \worse{3.9738}{5.53\%} \\
\textbf{w/o Sampling} &
\worse{0.3536}{14.47\%} & \worse{0.1885}{6.32\%} & \worse{0.1710}{2.64\%} &
\worse{0.4482}{6.72\%} & \worse{0.5774}{9.68\%} & \worse{3.7782}{0.34\%} \\
\midrule

\multicolumn{7}{l}{\textsc{Reversal} (\textsc{Full Steps})} \\
\textbf{M$^2$Sim (Ours)} &
\textbf{0.5434} & \textbf{0.2127} & 0.1986 & 0.4533 & 0.6065 & \textbf{3.8384} \\
\textbf{w/o LCT-State} &
\worse{1.1748}{116.19\%} & \worse{0.2536}{19.23\%} & \worse{0.2319}{16.77\%} &
\better{\textbf{0.4649}}{2.56\%} & \better{\textbf{0.6232}}{2.75\%} & \worse{4.0610}{5.80\%} \\
\textbf{w/o LCT-Policy} &
\worse{0.9359}{72.23\%} & \worse{0.2396}{12.65\%} & \better{\textbf{0.1808}}{8.96\%} &
\worse{0.4309}{4.94\%} & \worse{0.5592}{7.80\%} & \worse{4.2301}{10.20\%} \\
\textbf{w/o Sampling} &
\worse{0.7634}{40.49\%} & \worse{0.2281}{7.24\%} & \better{0.1897}{4.48\%} &
\worse{0.4214}{7.04\%} & \worse{0.5721}{5.67\%} & \worse{4.0404}{5.26\%} \\

\bottomrule
\end{tabular}
\label{tab:ablation_lct}
\end{table*}

\definecolor{CMTours}{HTML}{EA483F}
\definecolor{CMTmf}{HTML}{5D91BF}
\definecolor{CMTsr}{HTML}{6EB160}
\definecolor{CMTinst}{HTML}{FF983D}

\subsection{Ablation Study}
We ablate \textbf{LCT-State}, \textbf{LCT-Policy}, and \textbf{Sampling} to assess their respective contributions across \textsc{short-horizon}, \textsc{long-horizon}, and \textsc{reversal} simulations (Table~\ref{tab:ablation_lct}).

\textbf{(1) LCT-State is critical for long-horizon distribution tracking.}
Removing LCT-State yields the largest degradation on distributional metrics as the horizon extends.
In \textsc{long-horizon} simulation, \textbf{w/o LCT-State} sharply worsens KL ($0.3089 \rightarrow 0.8770$) and DTW ($0.1666 \rightarrow 0.2104$), indicating compounding errors in multi-step macro rollouts.
Similarly, in \textsc{reversal} simulation KL rises markedly ($0.5434 \rightarrow 1.1748$), showing that state-consistent mean-field evolution is crucial for non-monotonic dynamics.

\textbf{(2) Sampling in LCT-Policy is the main driver of F1.}
Ablating LCT-Policy degrades classification accuracy, especially under long horizons.
In \textsc{long-horizon} simulation, \textbf{w/o LCT-Policy} reduces Macro-F1 ($0.4805 \rightarrow 0.4314$) and Micro-F1 ($0.6393 \rightarrow 0.5765$), and in \textsc{Reversal} it further lowers Micro-F1 ($0.6065 \rightarrow 0.5592$).
Notably, removing \textbf{Sampling} alone causes a comparable F1 drop (Micro-F1 $0.6393 \rightarrow 0.5774$ in \textsc{long-horizon}, $0.6065 \rightarrow 0.5721$ in \textsc{Reversal}), indicating that sampling-based long-horizon selection drives most of the F1 gains.

\textbf{(3) Complementarity and a short-horizon trade-off.}
Overall, M$^2$Sim is the most balanced across settings, combining state-level rollout consistency with sampling-based long-horizon action selection.
Under \textsc{short-horizon}, \textbf{w/o LCT-State} and \textbf{w/o LCT-Policy} can slightly improve distance metrics (e.g., DTW), since short rollouts favor local curve fitting, while LCT constraints mainly benefit long-horizon stability and reversals.
Notably, \textbf{w/o LCT-State} slightly improves \textsc{Reversal} F1 (Macro-F1 $0.4533 \rightarrow 0.4649$, Micro-F1 $0.6065 \rightarrow 0.6232$) despite worse tracking, indicating per-step label fitting at the expense of coherent multi-step macro evolution.

\begin{figure*}[t]
    \centering
    \includegraphics[width=\linewidth, height=0.45\linewidth]{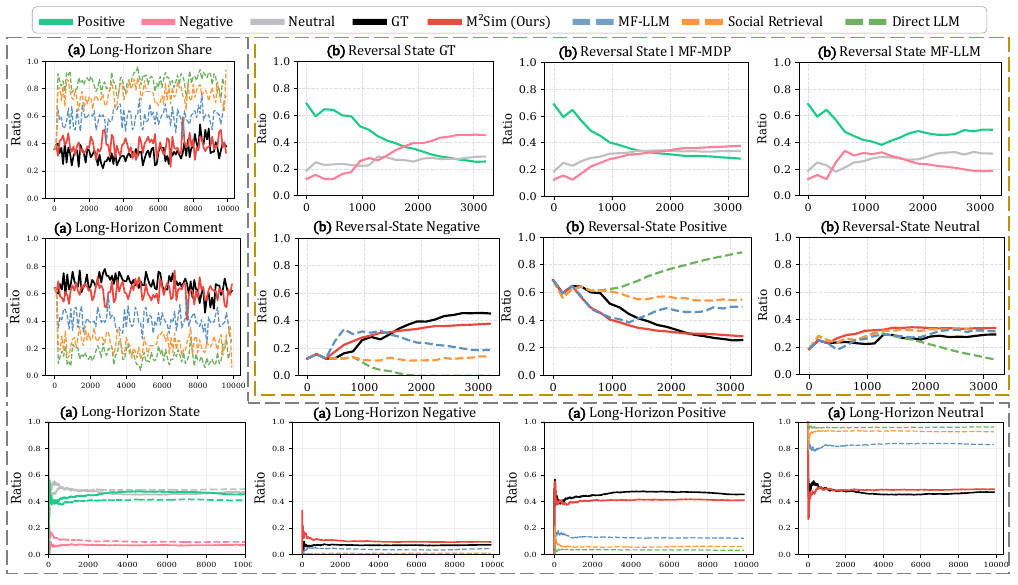}
    \caption{
    \textbf{Trajectory alignment of collective actions and state distributions.}
    We compare \textbf{Real data} with \textcolor{CMTours}{\textbf{M$^2$Sim (Ours)}}, \textcolor{CMTmf}{\textit{MF-LLM}}, \textcolor{CMTsr}{\textit{Social Retrieval}}, and \textcolor{CMTinst}{\textit{Direct LLM}}.
    (a) In long-horizon simulation, \textbf{M$^2$Sim} better matches action selection (e.g., comment and repost shares) and yields a closer fit to the state-distribution trajectory, while baselines drift more.
    (b) In a reversal event, \textbf{M$^2$Sim} captures the macro-level state reversal, whereas \textit{MF-LLM} fails to reproduce the turning point.
    }
    \label{fig3:analysis}
\end{figure*}

\subsection{Analysis and Discussion}
We analyze why M$^2$Sim improves long-horizon stability and reversal fidelity through a key limitation of \textit{MF-LLM}: its macro signal is \emph{implicit and narrative-smooth}, preserving coherence but failing to reliably control the \emph{amount} and \emph{timing} of state mass that drives micro actions. M$^2$Sim addresses this by representing the macro state as an \emph{explicit distribution} and feeding it into agent inputs, translating macro shifts into consistent micro reallocation. Additional analyses are provided in Appendix~\ref{appD.2:full analysis}.

\noindent\textbf{Long-horizon simulation.}
As shown in Fig.~\ref{fig3:analysis}a, \textit{MF-LLM}'s textual synopsis may describe ``the crowd is cooling down'' or ``debate is intensifying'' without specifying the mass of \emph{Neutral} versus \emph{Positive/Negative} states.
Different interpretations yield different action mixtures, with deviations compounding over long horizons.
Baselines tend toward a conservative regime with high neutrality, increasing repost share and distorting comment share.
Although \textit{MF-LLM} reduces drift relative to \textit{Direct LLM} and \textit{Social Retrieval}, it lacks precise control over neutrality.
M$^2$Sim predicts an explicit macro distribution and injects it into each agent's conditioning state: lower Neutral mass reduces neutral tendency and repost propensity, reallocating probability toward commenting and bringing action shares closer to real trajectories.
Training against future distributions keeps the macro trajectory aligned, limiting drift that a plausible textual synopsis alone may conceal.

\noindent\textbf{Reversal simulation.}
As shown in Fig.~\ref{fig3:analysis}b, reversal requires the macro signal to \emph{change direction} and propagate to micro decisions with the correct delay.
\textit{MF-LLM}'s temporally smooth, lexically consistent synopsis often remains ``moderate'' despite fluctuations, failing to express a decisive shift across state boundaries.
The micro policy consequently follows the earlier storyline, producing inertia and missing the turning point.
M$^2$Sim's distributional state can represent sharp redistribution, such as Neutral collapsing and Negative rising, without a gradual narrative bridge; agents directly observe this distribution, allowing the action mixture to pivot accordingly.
Long-horizon consistency training further downweights candidates whose rollouts remain ``middle-of-the-road'' when the real future requires reversal and upweights those producing the correct non-monotonic trajectory.
Together, these mechanisms reproduce the macro-level \emph{direction change} and subsequent micro-level \emph{behavioral switch} with correct temporal alignment.





\section{Conclusion}
We study large-scale social network simulation through the coupling of micro-level behaviors and macro-level collective dynamics.
Prior two-LLM mean-field simulators rely on macro-only updates, obscuring switching readiness and favoring self-reinforcing trajectories that miss opinion reversals.
We propose \textbf{M$^2$Sim}, a stateful mean-field simulator that couples latent opinion states of current participants with an evolving macro distribution through a learned state transition model.
Long-horizon trajectory agreement guides action reselection, linking current decisions to future collective dynamics.
Across real-world events, M$^2$Sim improves short-horizon fidelity, long-horizon stability, and reversal tracking while reducing simulation drift.

\subsection*{AI use statement}

We used generative AI tools to assist with language editing, translation, and checking the clarity and consistency of mathematical notation in the manuscript. We also used LLMs as components of the simulation framework and for semantic annotation in evaluation, as described in the methodology and experimental settings. The authors reviewed and revised the AI-assisted text and mathematical expressions to ensure that they accurately reflected the intended methodology. We take responsibility for the final content of this work, including its methods, claims, results, and AI-assisted content.

\subsection*{Ethics statement}

This work studies opinion dynamics in social networks to support understanding of collective behavior and evaluation of interventions in simulated settings. These techniques could also be misused for targeted persuasion, misinformation amplification, or manipulation at scale. We frame M$^2$Sim as a research and analytical tool and encourage applications with appropriate ethical review, transparency, and privacy safeguards. Simulated opinion states should not be interpreted as verified psychological attributes of real individuals. Moreover, LLM-generated behaviors and semantic annotations may contain biases or errors that affect the simulated trajectories and their evaluation. Although we evaluate predictions using explicit semantic dimensions and quantitative trajectory metrics, these measures do not eliminate annotation uncertainty or establish the effectiveness of interventions in real-world settings.

\bibliographystyle{ACM-Reference-Format}
\bibliography{main}

\newpage
\appendix
\onecolumn

\begin{center}
{\Large \textbf{Appendix}}
\end{center}

\noindent\textbf{A.} \hyperref[app:A]{Related Work}\par
\noindent\textbf{B.} \hyperref[app:B]{Theoretical Proofs}\par
\noindent\textbf{C.} \hyperref[app:C]{Detailed Experimental Setup}\par
\noindent\textbf{D.} \hyperref[app:D]{Additional Experimental Results and Analysis}\par

\section{Related Work}
\label{app:A}

\subsection*{Social Simulation Systems: Foundations and Limitations}
Traditional social simulation systems can be broadly classified into three paradigms.
\textit{(1) Mechanistic models} describe collective behavior through explicit equations or procedural dynamics such as discrete-event and system-dynamics~\citep{dynamics-lupeng-2021swarm,traditional-simulation-system-dynamics,traditional-simulation-discrete-events}.
\textit{(2) Empirical and statistical models} identify diffusion regularities from data, including PSP~\citep{PSP-2018} and peak-based participation dynamics~\citep{peak-height-lupeng-2018predicting,shapes-lupeng2019strength}.
\textit{(3) Agent-based models} capture emergent phenomena from local interactions among heterogeneous agents~\citep{Old-Agent-Based-Social-Simulation-2002,first-large-scale-agent-model-1996,squazzoni2008micro,Multi-Agent-Systems-application-2018}, with applications to collective dynamics~\citep{crowd-dynamics-2000-Nature}, market simulations~\citep{market-simulations-2006agent}, ecosystems~\citep{ecosystems-2005-Science}, and public policy~\citep{public-policy2000agents, Taxai-2023-mi}.
Despite their success, these paradigms often rely on handcrafted rules, simplified assumptions, and fixed parameters, which limits adaptability and makes robust long-horizon simulation difficult in open-ended, evolving settings.
Recent advances in large language models (LLMs)~\citep{LPT-2026-ACL} offer a complementary direction: replacing brittle rule templates with contextual, generative decision-making to better generalize while preserving rich semantic interactions.

\subsection*{LLM-Based Agent Social Simulation: Progress and Gaps}
Recent advances in LLMs~\citep{gpt-4-2023,Deepseek-r12025deepseek,DeepSeek-OCR2025deepseek,Qwen2.5-2025qwen2} have enabled cognitively enriched social simulations where agents communicate, reason, and adapt through natural language~\citep{COT-2022chain,logicagent-2026-ACL,Prompt-Design-2025prompt}.
At a high level, existing LLM-driven agent simulators fall into two paradigms.
\textit{(1) Prompted role-play ABM.}
These systems drive open-ended interactions mainly through role specifications and prompting (often with lightweight memory), and are typically used in small-scale or qualitative settings; examples include \textit{Generative Agents}~\citep{Stanford-town-2023,Stanford1000agents-2024} and \textit{S\textsuperscript{3}}~\citep{s3-2023}.
\textit{(2) Memory / retrieval-augmented ABM.}
These systems inject social context via heuristic retrieval or summarization (e.g., recent/popular responses) to improve scalability while retaining language-based decision making; examples include \textit{SocioVerse}~\citep{Socioverse2025socioverse}, \textit{AgentSociety}~\citep{Agentsociety-2025agentsociety}, and \textit{GA-S\textsuperscript{3}}~\citep{GAS32025-ga}.
Despite steady progress, most approaches remain \emph{micro-centric} and depend on prompting or heuristic memory/retrieval to carry evolving context, which is often brittle and struggles to retain decision-critical information over long horizons, resulting in unstable temporal dynamics and weak quantitative alignment with real-world collective trends.

\subsection*{Mean Field Approximation: Motivation and Challenges}
Mean field approximation~\citep{mf-approximation1998theory,mf-approximation2017refined} scales large multi-agent systems by replacing expensive pairwise interactions with interactions between each agent and a shared macro signal. This idea is formalized in mean-field game (MFG) theory~\citep{MFG-2007-JOM,MF-RL-2018-ICML}, where individual decisions and aggregate dynamics are coupled through a compact mean-field representation, and has been applied to domains such as social influence~\citep{MFG-2017-population-behavior,socialMFG-2016opinion}, traffic control~\citep{MFG-traffic-2024survey}, energy optimization~\citep{MFG-energy2012electrical}, and economic policy~\citep{ecosystems-2005-Science,MFG-economic2024mi}.
Despite this scalability, classical MFGs assume stylized behaviors and environment models, making it difficult to capture contextual, language-mediated decisions in realistic social settings; neural mean-field variants improve expressiveness~\citep{MFG-RL-2022-scalable-ICML} but remain limited when interactions are open-ended and semantics-rich. Recent LLM-based simulators introduce a shared macro signal~\citep{MF-LLM2025mf,PopSim-2025-liuwu}, yet macro-micro coupling remains weak: \textit{MF-LLM} lacks explicit latent micro states, while \textit{PopSim} relies on prompt-only design.
The key challenge is to learn a compact, decision-critical mean field that supports stateful decisions and long-horizon macro evolution without drifting from real trajectories.

\section{Theoretical Proofs}
\subsection{Derivation of Prop.2.1}
\begin{proof}
Fix a mean-field distribution $m$ and let
$s^\star \in \arg\max_{s\in\mathcal S} m(s)$.
By the law of iterated expectation and Assumption~(1),
\begin{align}
\mathbb{E}[m_{t+1}(s^\star)\mid m_t=m]
&=
\mathbb{E}_{\mathbf a_t\mid m_t=m}
\left[
\mathbb{E}[m_{t+1}(s^\star)
\mid m_t=m,\mathbf a_t]
\right] \\
&\ge
\mathbb{E}_{\mathbf a_t\mid m_t=m}
\left[
\widehat q(\mathbf a_t)(s^\star)-\epsilon
\right] \\
&=
\mathbb{E}[
\widehat q(\mathbf a_t)(s^\star)
\mid m_t=m]
-\epsilon.
\end{align}
Applying Assumption~(2) yields
\[
\mathbb{E}[m_{t+1}(s^\star)\mid m_t=m]
\ge m(s^\star)+\eta-\epsilon.
\]
Therefore,
\[
\delta_m(s^\star)
=
\mathbb{E}[m_{t+1}(s^\star)\mid m_t=m]
-m(s^\star)
\ge \eta-\epsilon.
\]
In particular, if $\eta\ge\epsilon$, then
$\delta_m(s^\star)\ge 0$.
\end{proof}

\subsection{Derivation of the Policy Model}
\label{app:B}

The policy model $\pi_{\phi}$ governs microscopic decision-making. We use a factorized policy over the active set:
\begin{equation}
\pi_{\phi}(\vec{a}_t \mid \vec{z}_t)
\;=\;
\prod_{i=1}^{n_t}\pi_{\phi}(a_{i,t}\mid z_{i,t}).
\label{eq:policy_factorized}
\end{equation}

\paragraph{Lookahead re-selection as latent-policy optimization.}
Starting from
\begin{equation}
\max_{\phi} \mathbb{E}_{\vec{a}\sim\pi_\phi}\!\left[R(\vec{a})\right],
\label{equ:RL_target}
\end{equation}
introduce a latent variable $\lambda\sim q(\lambda)$ and the conditional policy $\pi_\phi(\vec{a}\mid\lambda)$:
\begin{equation}
\begin{aligned}
\max_{\phi}\mathbb{E}_{\vec{a}\sim\pi_\phi}\!\left[R(\vec{a})\right]
&= \int R(\vec{a}) \pi_\phi(\vec{a})\, d\vec{a} \\
&=\int_{\vec{a}} R(\vec{a}) \int_{\lambda}\pi_\phi(\vec{a}\mid\lambda)q(\lambda)\,d\lambda \, d\vec{a} \\
&=\int_\lambda \Bigl(\int_{\vec{a}} R(\vec{a})\pi_\phi(\vec{a}\mid\lambda)\,d\vec{a}\Bigr) q(\lambda)\,d\lambda \\
&=\mathbb{E}_{\lambda \sim q(\lambda)} \Bigl[ \mathbb{E}_{\vec{a}\sim\pi_\phi(\cdot\mid\lambda)}[R(\vec{a})]\Bigr] \\
&=\mathbb{E}_{\lambda \sim q(\lambda)} \bigl[ \hat{R}_\phi(\lambda)\bigr],
\end{aligned}
\end{equation}
where
\begin{equation}
\hat{R}_{\phi}(\lambda) := \mathbb{E}_{\vec{a}\sim\pi_\phi(\cdot\mid\lambda)}[R(\vec{a}, \lambda)].
\end{equation}

To keep $q(\lambda)$ tractable while still preferring high-reward latent instances, we regularize it toward a trivial reference $p_{\rm trivial}$ (e.g., uniform over dropout masks), yielding a bi-level objective:
\begin{equation}
\max_{\phi} \max_{q} \mathbb{E}_{\lambda \sim q(\lambda)} [ \hat{R}_{\phi}(\vec{a}, \lambda)] - \frac{1}{\beta} \rm{KL} (q||p_{\rm trivial}).
\end{equation}
The inner maximization has the textbook solution
\begin{equation}
q^* \propto p_{\rm trivial}(\lambda) \exp(\beta \hat{R}_{\phi}(\vec{a}, \lambda)).
\end{equation}
Plugging $q^*$ back produces a log-sum-exp form (we keep the original notation; $Z(\phi)$ is the corresponding normalizer):
\begin{equation}
\begin{aligned}
&\max_{\phi} \mathbb{E}_{\lambda \sim q^* (\lambda)} [ \hat{R}_{\phi}(\vec{a}, \lambda)]- \frac{1}{\beta} \rm{KL} (q^*||p_{\rm trivial})\\
=& \max_{\phi} \mathbb{E}_{q^*}[ \hat{R}_{\phi}(\vec{a}, \lambda)] - \frac{1}{\beta}(\beta\mathbb{E}_{q^*}[ \hat{R}_{\phi}(\vec{a}, \lambda)]-\log^{Z(\phi)}) \\
=& \max_{\phi}  \frac{1}{\beta}\log^{\mathbb{E}_{p_{\rm trivial}}[\exp(\beta \hat{R}_{\phi}(\vec{a}, \lambda))]}.
\end{aligned}
\end{equation}
\begin{equation}
Z(\phi) = \int p_{\rm trivial}(\lambda)\exp(\beta\hat{R})d\lambda
=\mathbb{E}_{p_{\rm trivial}}[(\exp\beta\hat{R})].
\end{equation}

\paragraph{From maximization to a weighted cost.}
Rewriting the maximization as minimization with $V=-\hat{R}$ gives
\begin{align*}
\min_\phi  \frac{1}{\beta}\log^{\mathbb{E}_{p_{\rm trivial}}[\exp(\beta V_{\phi}(\vec{a}, \lambda))]}.
\end{align*}
Taking gradients reveals that the objective induces a \emph{softmax weighting} over $\lambda$:
\begin{align*}
\nabla_{\phi}\frac{1}{\beta}\log^{\mathbb{E}_{p_{\rm trivial}}[\exp(\beta V_{\phi}(\vec{a}^*, \lambda))]}
&= \frac{1}{\beta}\frac{1}{Z(\phi)} \nabla_\phi Z(\phi) \\
&= \frac{1}{\beta}\frac{1}{Z(\phi)} \mathbb{E}_{p}\!\left[\beta\exp(\beta V_\phi)\nabla_\phi V_\phi\right] \\
&= \frac{1}{Z(\phi)} \mathbb{E}_{p}\!\left[\exp(\beta V_\phi)\nabla_\phi V_\phi\right] \\
&= \mathbb{E}_{p}\!\left[\frac{\exp(\beta V_\phi)}{Z(\phi)}\nabla_\phi V_\phi\right].
\end{align*}
and integrating yields
\begin{equation}
\frac{1}{\beta}\log^{\mathbb{E}_{p_{\rm trivial}}[\exp(\beta V_{\phi}(\vec{a}^*, \lambda))]}=\mathbb{E}_{p_{\rm trivial}}[w V_\phi],
\end{equation}
where
\[
w(\vec{a}, \lambda)=\rm \frac{\exp(\beta V_\phi)}{Z(\phi)}
\]
is exactly a $\rm softmax$ over $\lambda$ with temperature $\beta$. Therefore the final objective becomes
\begin{equation}
\min_\phi \mathbb{E}_{\lambda \sim p_{\rm trivial}}[w V_\phi(\vec{a}, \lambda)].
\end{equation}

\textbf{Instantiating the long-horizon cost.}
We define the long-horizon cost as a discounted divergence:
\begin{equation}
V_\phi\!\left(\vec{a}, \lambda\right)
=
\sum_{k=1}^{K} \gamma^{k-1}
\mathrm{KL}\!\left(m_{t+k}^{\star}\,\|\,\hat m_{t+k}(\pi_\phi(\vec{a}), \lambda)\right).
\end{equation}
Aggregating over simulation time $t$ and active agents $n_t$ yields the prediction loss
\begin{align}
\mathcal{L}_{\rm pred} &= \sum_{t=1}^{T}\sum_{i=1}^{n_t}\mathbb{E}_{p}[w_{i,t}\sum_{k=1}^{K} \gamma^{k-1} 
\mathrm{KL}\!\left(m_{t+k}^{\star}\,\|\,\hat m_{t+k}(\pi_\phi)\right)].
\end{align}

To stabilize optimization, we include an auxiliary text-supervision term (ground-truth actions) to reduce variance:
\begin{equation}
\label{eq:policy_text}
\mathcal{L}_{\mathrm{text}}(\phi)
=
-\mathbb{E}\!\left[\sum_{t=1}^{T}\sum_{i=1}^{n_t}
\log \pi_{\phi}\!\bigl(\vec{a}_{i,t}^{\star}\mid \lambda\bigr)\right].
\end{equation}
The following bound shows that increasing $\pi_\phi(\vec{a}^*|\lambda)$ tightens the variance of the induced cost.
\begin{theorem}
If $\vec{a}$ is a discrete random vector and $|V_\phi| \le M$, then $\mathrm{Var}(V) \le 4M^2(1-\pi_\phi(\vec{a}^*|\lambda))^2$.
\end{theorem}
\begin{align*}
\left|\hat{R}_\phi(\vec{a}, \lambda) - R(\vec{a}^*, \lambda)\right|
&= \left|\sum_{\vec{a}} R_\phi(\vec{a}, \lambda)\pi_\phi(\vec{a}\mid\lambda) - R(\vec{a}^*, \lambda)\right| \\
&\le \sum_{\vec{a}} \pi_\phi(\vec{a}\mid\lambda)\left|R_\phi(\vec{a}, \lambda) - R(\vec{a}^*, \lambda)\right| \\
&\le 2M \sum_{\vec{a}\ne \vec{a}^*} \pi_\phi(\vec{a}\mid\lambda) \\
&= 2M \left(1-\pi_\phi(\vec{a}^*\mid\lambda)\right).
\end{align*}
\begin{equation}
\mathrm{Var}(\hat{R})
= \mathbb{E}\!\left[(\hat{R}-\mathbb{E}[\hat{R}])^2\right]
\le \mathbb{E}\!\left[(\hat{R}-R(\vec{a}^{*},\lambda))^2\right]
\le 4M^2\bigl(1-\pi_{\phi}(\vec{a}^{*}\mid\lambda)\bigr)^2.
\end{equation}

Finally, we optimize $\pi_{\phi}$ via the two-term objective
\begin{equation}
\label{eq:policy_total}
\mathcal{L}_{\mathrm{total}}(\phi)
=\mathcal{L}_{\mathrm{pred}}(\phi) +
\alpha\,\mathcal{L}_{\mathrm{text}}(\phi).
\end{equation}
In practice, $\lambda$ is instantiated by in-place dropout sampling within the forward pass, so drawing $\lambda\sim p_{\rm trivial}$ corresponds to sampling dropout subnetworks, while the weights $w$ implement a soft selection over these dropout policy instances.

\section{Detailed Experimental Setup}
\label{app:C}

\subsection{Training Curves and Hyperparameters}
\label{app:C.1 training}
\begin{table*}[t]
\centering
\footnotesize
\setlength{\tabcolsep}{10pt}      
\renewcommand{\arraystretch}{0.98}  
\caption{Model and training hyperparameters. The left block reports the state transition model (Event Transformer) used in M$^2$Sim; the right block reports LLM-based components fine-tuned from Qwen2-1.5B-Instruct.}
\label{tab:all_hparams}
\begin{tabular}{@{}l c | l c c c@{}}
\toprule
\multicolumn{2}{c|}{\textbf{State Transition (Event Transformer)}} &
\textbf{Hyperparameter} & \textbf{Mean Field} & \textbf{IB-Tune} & \textbf{LCT-Tune} \\
\midrule
Hidden size $d_{\text{model}}$ & 256 &
Base model & Qwen2-1.5B & Qwen2-1.5B & Qwen2-1.5B \\
\#layers / \#heads & 3 / 8 &
Max sequence length & 2048 & 2048 & 2048 \\
Max sequence length & 4096 &
Training dataset & WEIBO & WEIBO & WEIBO \\
Dropout & 0.1 &
Training batch size & 256 & 256 & 256 \\
FFN dimension $d_{ff}$ & 1024 &
Micro batch size & 8 & 8 & 8 \\
Text Encoder & BERT &
Max epochs & 1 & 1 & 1 \\
Optimizer & AdamW &
Random Seed & 46 & 46 & 46 \\
Learning rate & $2\times10^{-5}$ &
Learning rate & $5\times10^{-7}$ & $5\times10^{-7}$ & $5\times10^{-7}$ \\
Batch size (events) & 4 &
LoRA rank/alpha & 64/64 & 64/64 & 64/64 \\
Max epochs & 20 &
Prediction weight $\alpha_{\text{coeff}}$ & -- & -- & 0.5 \\
Weight decay & $1\times10^{-5}$ &
Lookahead horizon $K$ & -- & -- & 30 \\
Gradient clip & 1.0 &
\#candidates $J$ & -- & -- & 4 \\
Loss function & $\mathcal{L}_{\mathrm{trans}}$ &
Loss function & $\mathcal{L}_{\text{mean-field}}$ & $\mathcal{L}_{\text{policy}}$ & $\mathcal{L}_{\text{LCT}}$ \\

\bottomrule
\end{tabular}
\end{table*}

\begin{figure*}[t]
    \centering
    \includegraphics[width=\linewidth]{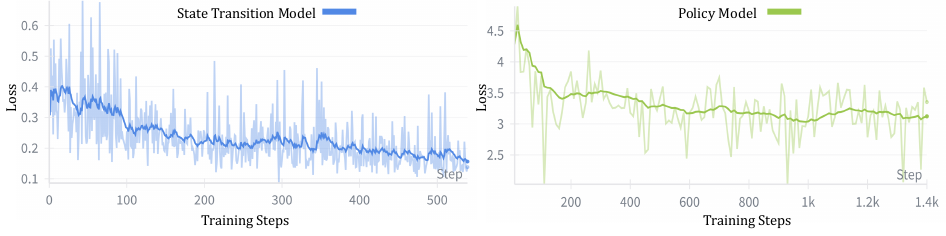}
    \caption{\textbf{Training loss curves for M$^2$Sim.}
    \textbf{Left:} Event Transformer (state transition model) optimized by $\mathcal{L}_{\mathrm{trans}}$ (KL divergence between predicted and empirical mean-field distributions).
    \textbf{Right:} Policy model optimized by $\mathcal{L}_{\mathrm{total}}=\mathcal{L}_{\mathrm{pred}}+\alpha\,\mathcal{L}_{\mathrm{text}}$, combining discounted $K$-step rollout discrepancy with action NLL supervision.}
    \label{fig4:loss}
\end{figure*}

\noindent\textbf{Summary.}
All experiments are run on two Tesla V100S-PCIE-32GB GPUs.
Table~\ref{tab:all_hparams} reports the hyperparameters for both the state transition model (Event Transformer) and all LLM components (Mean Field IB-Tune, Policy IB-Tune, and Policy LCT-Tune), with LLMs fine-tuned from \texttt{Qwen2-1.5B-Instruct} using LoRA on WEIBO under a unified data format and sequence length.

\noindent\textbf{Training curves.}
Figure~\ref{fig4:loss} shows the optimization dynamics of our two core modules.
The \textit{state transition model} is trained with the KL-based transition loss $\mathcal{L}_{\mathrm{trans}}$ against the empirical mean-field distribution, while the \textit{policy model} is trained with $\mathcal{L}_{\mathrm{total}}=\mathcal{L}_{\mathrm{pred}}+\alpha\,\mathcal{L}_{\mathrm{text}}$, combining discounted $K$-step rollout divergence and NLL supervision on ground-truth actions.

\subsection{Detailed Settings}
\label{appC.2:detailed setup}

\noindent\textbf{Dataset.}
We follow \textit{MF-LLM}~\citep{MF-LLM2025mf} and adopt the WEIBO corpus~\citep{weibo-2016} as the primary benchmark, which contains $5{,}000+$ real-world events with temporally ordered individual responses and rich individual profiles, covering categories such as \emph{Crime, Culture, Health, News, Politics, Sports,} and \emph{Technology}.
To better stress-test \emph{reversal} and delayed-commitment dynamics beyond the original benchmark, we additionally curate a complementary \textbf{Reversal} collection by crawling public discussions from Weibo and Douyin, spanning domains including \emph{Education, Economy, Society, Environment,} and \emph{Campus}.
We convert all newly collected data into the same \textit{MF-LLM} event-centric format (event timeline, per-timestep active individual set, responses, and profiles), enabling plug-and-play training/evaluation under an identical simulation interface.
The Reversal set contains much longer trajectories than WEIBO (up to $40{,}000+$ timesteps), so we use full-length rollouts, as short horizons rarely exhibit reversals.
For WEIBO, we evaluate long events (length $>1000$) under both the default $300$-step rollout (\textit{MF-LLM}) and full-trajectory settings.

\definecolor{CMTours}{HTML}{EA483F}
\definecolor{CMTmf}{HTML}{5D91BF}
\definecolor{CMTsr}{HTML}{6EB160}
\definecolor{CMTinst}{HTML}{FF983D}

\begin{figure*}[t]
    \centering
    \includegraphics[width=\linewidth]{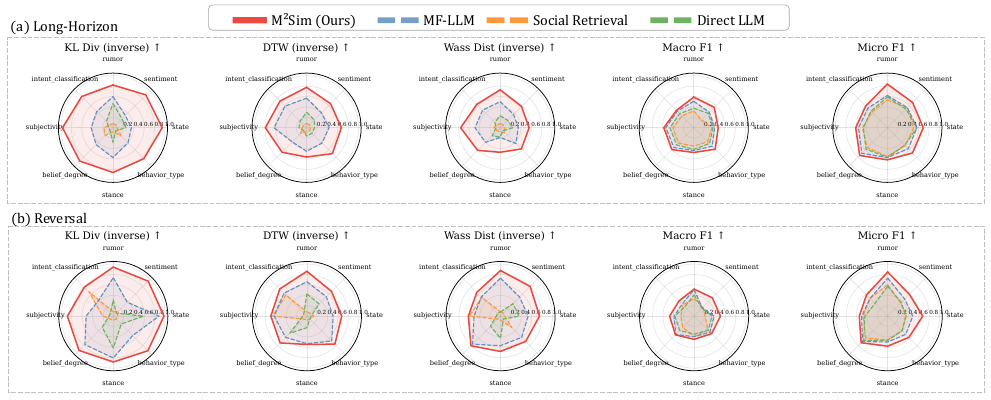}
    \caption{
\textbf{Radar plot} comparing \textcolor{CMTours}{\textbf{M$^2$Sim (Ours)}} and three baselines
(\textcolor{CMTmf}{\textbf{MF-LLM}},
\textcolor{CMTsr}{\textbf{Social Retrieval}},
\textcolor{CMTinst}{\textbf{Direct LLM}})—
on 5 distributional metrics and 8 semantic dimensions of actions.
KL Divergence (Inverse), Wasserstein Distance (Inverse), and DTW Distance (Inverse) are inversely normalized so larger values indicate better performance, alongside Macro F1 and Micro F1.
Larger areas denote superior performance, with \textcolor{CMTours}{\textbf{M$^2$Sim}} showing the highest semantic fidelity.
    }
    \label{fig5:radar}
\end{figure*}

\noindent\textbf{Evaluation Metrics.}
We adopt a micro-to-macro evaluation protocol:

\textbf{(i) Micro-level (individual actions).} We focus on \emph{semantic-related} evaluation by annotating each generated (and real) action into one of $8$ semantic dimensions using an LLM-based annotator (GPT-4o-mini). This LLM-based evaluation protocol follows prior mean-field simulators (\textit{MF-LLM}) and has been empirically validated therein.
\begin{itemize}[leftmargin=*]
    \item \textbf{Rumor.} Whether the action \emph{spreads} the discussed claim (believes/amplifies) or \emph{counters} it (questions/refutes/clarifies).
    \item \textbf{Sentiment.} The expressed emotional tone (including sarcasm/irony), e.g., angry, calm, happy, sad, fear, surprise.
    \item \textbf{State.} Overall polarity toward the topic: positive, negative, or neutral, including subtle negativity.
    \item \textbf{Behavior.} Interaction type: \texttt{share} (repost/forward) vs.\ \texttt{comment} (textual response).
    \item \textbf{Stance.} Position toward the topic: support, oppose, or neutral, including implicit opposition.
    \item \textbf{Belief.} Perceived truthfulness: believe vs.\ doubt (skepticism, requests for evidence, denial).
    \item \textbf{Subjectivity.} Subjective personal opinion vs.\ objective factual description.
    \item \textbf{Intent.} Communicative goal: question (seeking clarification), promotion (disseminating), or opinion (expressing viewpoint).
\end{itemize}

\textbf{(ii) Macro-level (distributional dynamics).}
Building on the micro-level annotations, at each timestep $t$ we map every action into a discrete label (under a chosen dimension) and aggregate the $n_t$ actions into an empirical categorical distribution $p_t$; we compute the same distribution $\hat p_t$ for generated actions. We then compare the real trajectory $\{p_t\}_{t=1}^{T}$ and the generated trajectory $\{\hat p_t\}_{t=1}^{T}$ using:
\begin{enumerate}[leftmargin=*]
    \item \textbf{KL Divergence.} $\mathrm{KL}(p_t\,\|\,\hat p_t)$, averaged over timesteps, penalizing mismatched probability mass and being sensitive to mode dropping.
    \item \textbf{Wasserstein Distance.} $W(p_t,\hat p_t)$, averaged over timesteps, measuring the cost of transporting probability mass and being more robust to small support shifts.
    \item \textbf{Dynamic Time Warping (DTW).} DTW between the two time series $\{p_t\}$ and $\{\hat p_t\}$ (or scalar projections per label), evaluating temporal alignment by allowing elastic matching across timesteps and penalizing phase shifts.
    \item \textbf{Negative Log-Likelihood (NLL).} The average $-\log \pi_{\phi}(a_{i,t}^{\star}\mid z_{i,t})$ over all ground-truth actions, measuring how well the learned policy assigns probability to real behaviors.
    \item \textbf{Macro-F1.} F1 computed from predicted vs.\ real labels and averaged across classes (treating each class equally), highlighting performance on minority labels.
    \item \textbf{Micro-F1.} F1 computed by aggregating true/false positives across all actions before forming F1, emphasizing overall accuracy dominated by frequent labels.
\end{enumerate}



\begin{figure*}[t]
    \centering
    \includegraphics[width=\linewidth]{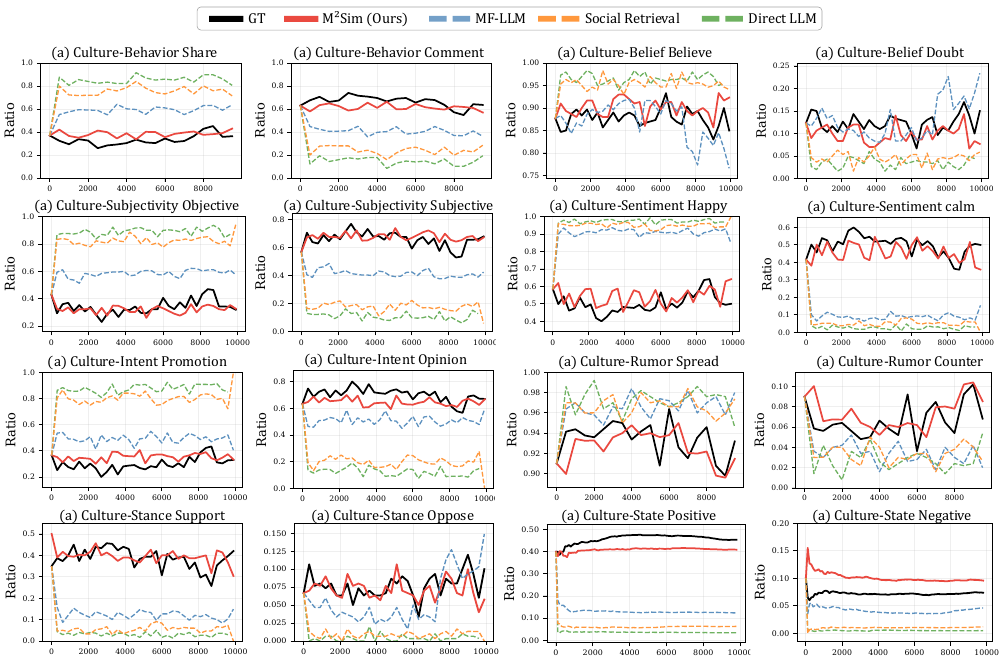}
    \caption{
    \textbf{Comparison on 8 semantic dimensions.}
We compare \textcolor{CMTours}{\textbf{M$^2$Sim (Ours)}}, \textcolor{CMTmf}{\textit{MF-LLM}}, \textcolor{CMTsr}{\textit{Social Retrieval}}, and \textcolor{CMTinst}{\textit{Direct LLM}}.
    }
    \label{fig7:8dimension}
\end{figure*}

\section{Additional Experimental Results and Analysis}
\label{app:D}

\subsection{Full Results}
\label{appD.1:full result}

\paragraph{Semantic fidelity under evolving states.}
Fig.~\ref{fig5:radar} evaluates semantic fidelity over eight semantic-related dimensions (excluding \textbf{NLL Loss}); for \textbf{KL}/\textbf{Wasserstein}/\textbf{DTW} we use inverse normalization so higher is better.
Across both \emph{long-horizon} and \emph{reversal} settings, M$^2$Sim consistently achieves the largest radar area.
Most notably, M$^2$Sim leads by a wide margin on the \textbf{State} axis, showing that it best preserves state-consistent action meaning—enabled by explicitly modeling and coupling the macro sentiment distribution (positive/neutral/negative) with each agent’s latent state during decision making.
This state grounding propagates to closely related dimensions.
Compared with all baselines, M$^2$Sim shows clear gains on \textbf{Sentiment} and \textbf{Stance}, indicating that expressed attitudes track the underlying state rather than drifting with surface text patterns.
We also observe a substantial improvement on \textbf{Behavior}, which is tightly linked to sentiment states (e.g., supportive vs.\ opposing engagement patterns under positive vs.\ negative shifts), consistent with M$^2$Sim’s better distributional alignment (\textbf{KL}/\textbf{Wass}/\textbf{DTW}) and higher \textbf{Macro F1}/\textbf{Micro F1}.
Overall, the radar shapes suggest that jointly modeling macro and micro states is key to producing semantically faithful collective actions, especially when trajectories must adapt over time.

\begin{table*}[h!]\footnotesize
    \centering
    \caption{Representative events with opinion reversal across domains.}
    \label{tab:event_cases}
    \renewcommand{\arraystretch}{1.15}
    \setlength{\tabcolsep}{8pt}
    \begin{tabular}{c p{2.4cm} p{1.6cm} p{5.5cm} p{4cm}}
    \toprule
    ID & Title & Domain & Description & Distinctive Features \\
    \midrule
    a &
    Celebrity Coordinated Posting Controversy &
    Culture &
    A public figure released a critical social media post containing an unintended scheduling artifact, revealing coordinated narrative behavior on online platforms. &
    Observable evidence of organized opinion coordination; ineffective denial; persistent credibility erosion. \\

    b &
    Global Mathematics Competition Eligibility Dispute &
    Education &
    An unexpected finalist from a non-traditional background initially triggered widespread admiration, later reversed by official findings of rule violations. &
    Rapid shift from emotional endorsement to scrutiny of procedural fairness and integrity. \\

    A &
    Full Registration-Based IPO Reform &
    Economy &
    The implementation of a registration-based IPO system, alongside market-stabilization measures, initially generated optimism but later faced skepticism as outcomes diverged from expectations. &
    Transition from policy-driven enthusiasm to institutional performance reassessment. \\

    B &
    Gradual Retirement Age Adjustment Policy &
    Society &
    A phased retirement age adjustment policy initially sparked strong resistance, later moderated by clarifications emphasizing flexibility and voluntariness. &
    Shift from collective anxiety to pragmatic individual adaptation. \\

    C &
    Nuclear Wastewater Discharge and Public Response &
    Environment &
    A cross-border environmental discharge plan provoked intense public anxiety, followed by a gradual shift toward long-term scientific monitoring frameworks. &
    Opinion evolution from acute panic to evidence-based risk oversight. \\

    D &
    University Library Harassment Allegation &
    Campus &
    An online harassment allegation prompted strong initial support for perceived victims, later complicated by additional evidence and procedural disclosures. &
    Rebalancing between moral advocacy and procedural objectivity. \\
    \bottomrule
    \end{tabular}
\end{table*}

\subsection{Full Analysis}
\label{appD.2:full analysis}

\textbf{Case Events.}
The case events used in Fig.~\ref{fig3:analysis} (Events a and b) and Fig.~\ref{fig6:state} (Events A-D) are summarized in Table~\ref{tab:event_cases}.

\paragraph{Long-horizon semantic alignment in full dynamic simulations.}
Fig.~\ref{fig7:8dimension} plots the 10,000-step trajectories of eight semantic dimensions for event (a) in the Great-Wall long-horizon simulation.
M$^2$Sim (red) is consistently the closest to GT (black) in both \emph{level} and \emph{trend}: across \textbf{Behavior} (Share/Comment), \textbf{Subjectivity} (Objective/Subjective), \textbf{Intent} (Promotion/Opinion), \textbf{Belief} (Believe/Doubt), and \textbf{Rumor} (Spread/Counter), it reproduces GT-like plateaus and fluctuations rather than drifting toward saturated, nearly constant curves.
\textit{MF-LLM} is generally second-best but shows noticeable offsets and occasional instability, while \textit{Social Retrieval} and \textit{Direct LLM} frequently collapse into extreme, unbalanced patterns.
The dominant gap appears on \textbf{State}: M$^2$Sim closely matches GT on both \textbf{State Positive} and \textbf{State Negative}, whereas all baselines underestimate them (curves near zero), effectively collapsing toward an overly neutral distribution.
This state mismatch propagates to state-adjacent semantics—\textbf{Sentiment} (Happy/Calm) and \textbf{Stance} (Support/Oppose)—where baselines tend to produce one-sided polarity that does not follow GT’s balance.
With explicit conditioning on the macro state distribution and each agent’s latent state, M$^2$Sim maintains coherent sentiment and stance dynamics, which further translates into more realistic \textbf{Behavior} trajectories over long horizons.

\paragraph{State reversals under exogenous signals in reversal events.}
Beyond the reversal case in Fig.~\ref{fig3:analysis}(b), we further evaluate four additional reversal events with horizons ranging from 6{,}000 to 40{,}000 steps.
These events feature long-run dynamics where the macro state distribution (positive/neutral/negative) can \emph{reversal} under exogenous signals.
As shown in Fig.~\ref{fig6:state}, M$^2$Sim consistently captures both the turning points and the post-reversal trends, staying close to GT across events rather than converging to a stationary trajectory.
This advantage comes from M$^2$Sim’s explicit \emph{state coupling}: it conditions decisions jointly on the macro state distribution and each agent’s latent state, while injecting the exogenous signal as a direct driver of state evolution.
When the exogenous signal shifts the macro distribution, M$^2$Sim propagates the change through agent states and back into the aggregate trajectory, producing the correct reversal dynamics.
In contrast, \textit{MF-LLM} relies on a text-based, coarse macro summary that becomes increasingly blurry over long horizons; after a short warm-up, it loses discriminative state information, so the predicted state curves remain nearly unchanged even when GT reverses.
This is further reflected in the fine-grained components: M$^2$Sim aligns best with GT on \textbf{Positive}, \textbf{Negative}, and \textbf{Neutral} simultaneously, whereas baselines typically under-react and collapse toward an overly neutral or biased mixture.

\begin{figure}[ht]
    \centering
    \includegraphics[width=\linewidth]{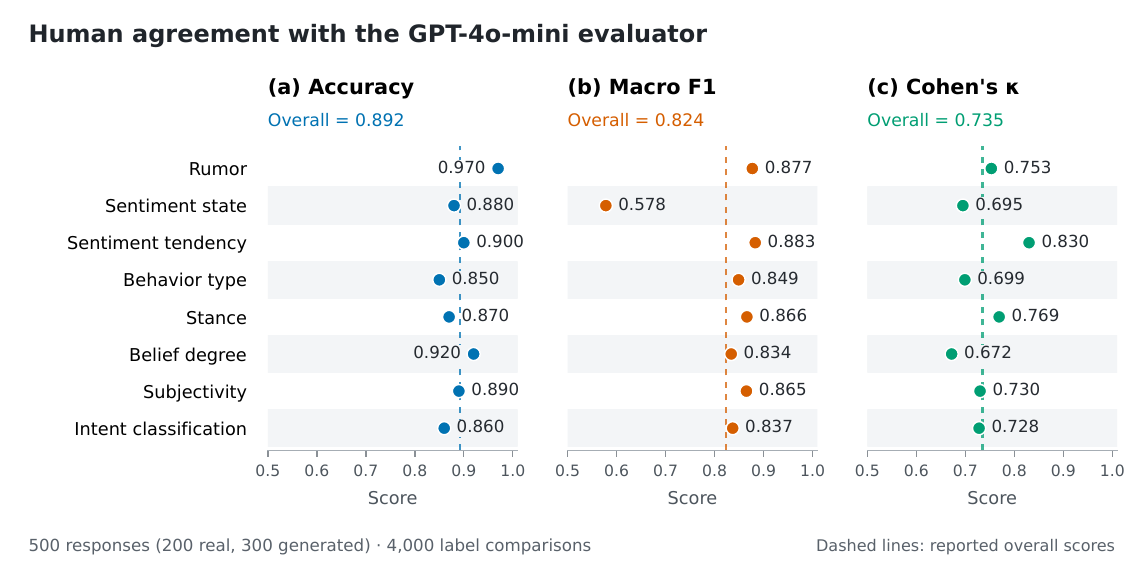}
    \caption{\textbf{Human validation of the GPT-4o-mini evaluator.}
    Human--evaluator agreement across eight semantic dimensions, measured by (a) Accuracy, (b) Macro F1, and (c) Cohen's $\kappa$.
    Dashed lines indicate the reported overall scores.}
    \label{fig:human_validation}
\end{figure}

\subsection{Human Validation of LLM-Based Evaluation}
\label{app:human_validation}

To assess the reliability of GPT-4o-mini as a semantic evaluator, we randomly sampled 500 responses, comprising 200 real and 300 generated responses.
The samples were evenly assigned to five human annotators, each labeling 100 responses without access to the evaluator's labels.
Each response was annotated along the same eight semantic dimensions used in our evaluation, yielding 4,000 human--evaluator comparisons.

Figure~\ref{fig:human_validation} shows overall Accuracy, Macro F1, and Cohen's $\kappa$ of 0.892, 0.824, and 0.735, respectively, supporting the use of GPT-4o-mini for semantic evaluation.
Agreement varies across dimensions, with sentiment state exhibiting a lower Macro F1 of 0.578, indicating that annotation uncertainty remains.

\begin{figure*}[t]
    \centering
    \includegraphics[width=\linewidth, height=1.3\linewidth]{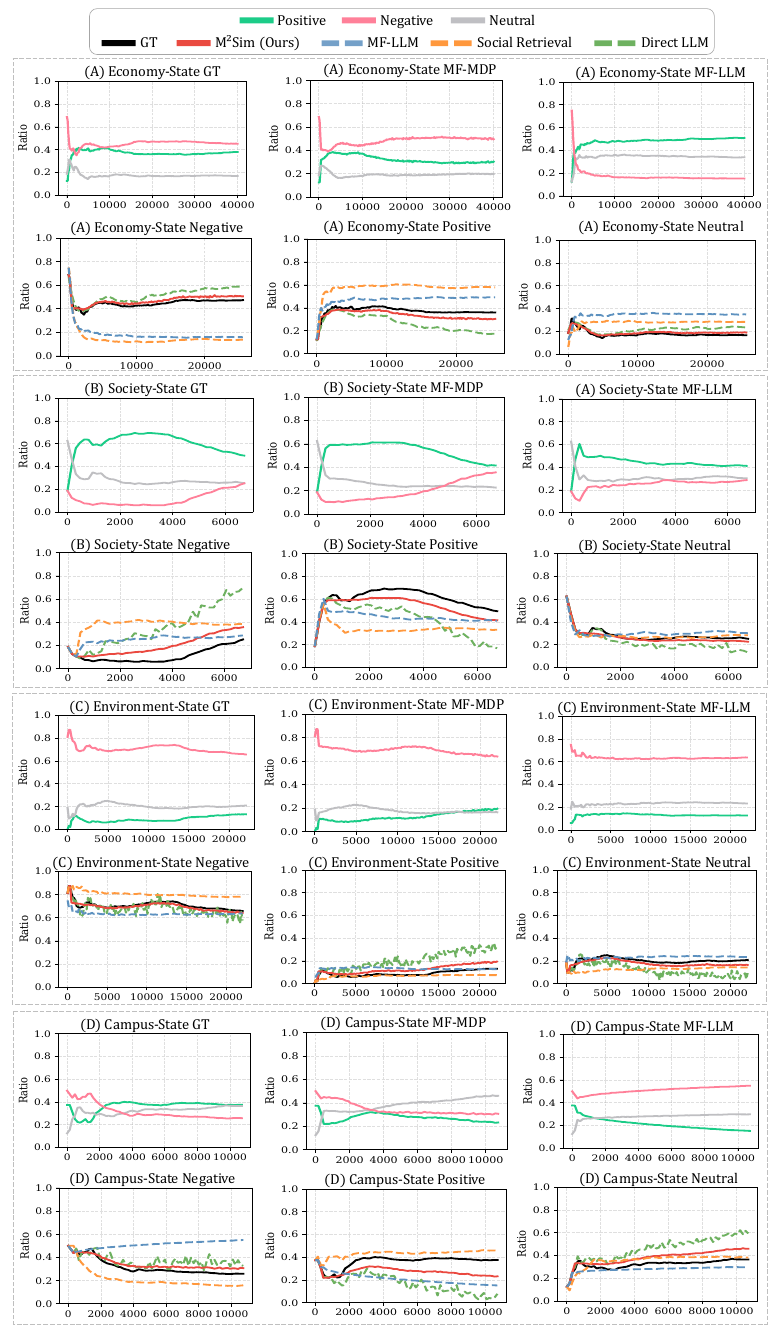}
    \caption{
    \textbf{State-distribution trajectories on four reversal events.}
We compare \textbf{Real data} with \textcolor{CMTours}{\textbf{M$^2$Sim (Ours)}}, \textcolor{CMTmf}{\textit{MF-LLM}}, \textcolor{CMTsr}{\textit{Social Retrieval}}, and \textcolor{CMTinst}{\textit{Direct LLM}}.
    }
    \label{fig6:state}
\end{figure*}

\end{document}